\documentclass[12pt]{article}
\usepackage{graphicx}
\begin{document}

\centerline {\bf \large Habitat width along a latitudinal gradient}

\bigskip

\centerline{D. Stauffer$^1$, C. Schulze$^1$, K. Rohde$^2$}

\centerline{$^1$ Theoretische Physik, Universit"at, D-50923 K\"oln, 
Euroland}

\centerline{$^2$ Zoology, University of New England, Armidale NSW 2351, Australia.}

\bigskip

Abstract:

\noindent
We use the Chowdhury ecosystem model, one of the most complex
agent-based ecological models, to test the latitude-niche breadth
hypothesis, with regard to habitat width, i.e., whether tropical species 
generally have narrower
habitats than high latitude ones. Application of the model has given
realistic results in previous studies on latitudinal gradients in
species diversity and Rapoport's rule. Here we show that tropical
species with sufficient vagility and time to spread into adjacent
habitats, tend to have wider habitats than high latitude ones,
contradicting the latitude-niche breadth hypothesis.

Keywords:

\noindent
Chowdhury ecosystem model, latitude-niche breadth hypothesis,
Rapoport's rule, latitudinal gradients in species diversity,
vagility, species-area relationship, fractal dimensions.

\section{Introduction}

According to a widely held view, an increase in diversity must result
in a narrowing of niches, in denser species packing. Thus, according
to Rosenzweig \& Ziv (1999) "Theory suggests that higher diversity
should shrink niches, allowing the coexistence of more species".
Applied to latitudinal gradients, the much greater species richness
in the tropics than in colder environments is thought possible only
because species are more densely packed, i.e., have smaller niches.
This view (the so called latitude-niche breadth hypothesis) can be
traced back to MacArthur (MacArthur 1965, 1969, 1972; MacArthur \&
Wilson 1967), but is probably even older.  There is some empirical
evidence for this view (e.g., MacArthur 1965, 1969; Moore 1972), and
much against it (e.g., Rohde 1980; Novotny \& Basset 2005). For
example, concerning one aspect of the niche, the latitudinal range of
a species, some studies have provided support for the view that
latitudinal ranges are narrower at low latitudes (Rapoport's rule,
e.g. Stevens 1989), whereas others have found no support or evidence
for an opposite trend (e.g., Rohde et al. 1993). Rohde (1998)
therefore suggested two opposing trends: newly evolved species with
little vagility may have narrower ranges in the tropics, species with
greater vagility and of sufficient age to spread into adjacent areas
may have larger ranges. The same may apply to habitat width and 
niche width in general.

In this paper, we use the Chowdhury ecosystem model (Chowdhury \&
Stauffer, 2005; Stauffer et al. 2005) to examine the latitude-niche
breadth hypothesis with regard to one of the most important niche dimensions,
the habitat width. We check the effect of vagility and age of ecosystem on
habitat width. We have applied the model before (Rohde \& Stauffer 2005;
Stauffer \& Rohde 2006) to study the variation of species diversity
and latitudinal ranges with latitude, comparing cold with tropical
regions in simulations of the whole range of latitudes in a lattice
model, and got realistic results.

\section{Methods}

The Chowdhury model (Chowdhury \& Stauffer, 2005; Stauffer et al.
2005) is one of the most complex agent-based (Billari et al. 2006)
ecological models (P\c ekalski, 2004; rimm \& Railsback, 2005) and has
been reviewed e.g. in Stauffer et al. (2006). Each species may move
to a neighbouring lattice site where it is still the same species.
Further details are given in an appendix.

One open question is the fractal dimension $D$ of the number $N$ of
species found in a square of side length $L$:
$$ N \propto L^D. $$
Empirically, fractal dimensions $ 1.2 \le D \le 2.3$ are given
by Rosenzweig (1995),
whereas $D=2$ would correspond to a trivial proportionality of the number of
species and the area in which they are counted. The rationale behind our
comparison of fractal dimensions is: in the extreme case, the largest square
could have a single species, which is also found in the smallest
square, i.e., the slope is 0, the species' habitat is very wide. On the
other hand, the largest square could have 100 species, 10 of which
are also found in the smallest square, i.e. the slope is much
steeper, the habitats are much narrower.

An earlier attempt (Stauffer and P\c ekalski (2005) roughly gave this simple
proportionality when it used the low vagility $d$ (diffusivity) which
gave good results in Rohde \& Stauffer (2005) and Stauffer \& Rohde
(2006). However, while in these papers we simulated the whole Earth
from the north pole to the south pole, tests of the above exponent $D$
should look at smaller, more homogenous regions. Thus, the
vagility $d$, which is the probability that a species invades a
neighbouring lattice site during one time step, has to be larger for
smaller lengths associated with the neighbor distance. Thus we now
use larger $d$ than in Stauffer \& P\c ekalski (2005), Rohde \&
Stauffer (2005) and Stauffer \& Rohde (2006) and also systematically
vary the observation time (measured in Monte Carlo steps per site; we
refrain from identifying it with years). We simulate (in most cases) ten $L 
\times L$ square lattices, with the other parameters besides vagility and
observation time as in Rohde \& Stauffer (2005) and Stauffer \& Rohde
(2006). Each such simulation either refers to tropical or to high
latitude (here referred to as polar) regions. As in Rohde \& Stauffer
(2005) and Stauffer \& Rohde (2006), we use the standard Chowdhury
model for the simulation of the tropical region, while for the
polar region the birth rate is reduced by a factor 4. This birth rate
is the probability per iteration that offspring reaching maturity is
being produced, due to the harsh living conditions in polar regions we
assume this probability to be four times lower than in the tropics.

\section{Results}

\begin{figure}[hbt]
\begin{center}
\includegraphics[angle=-90,scale=0.25]{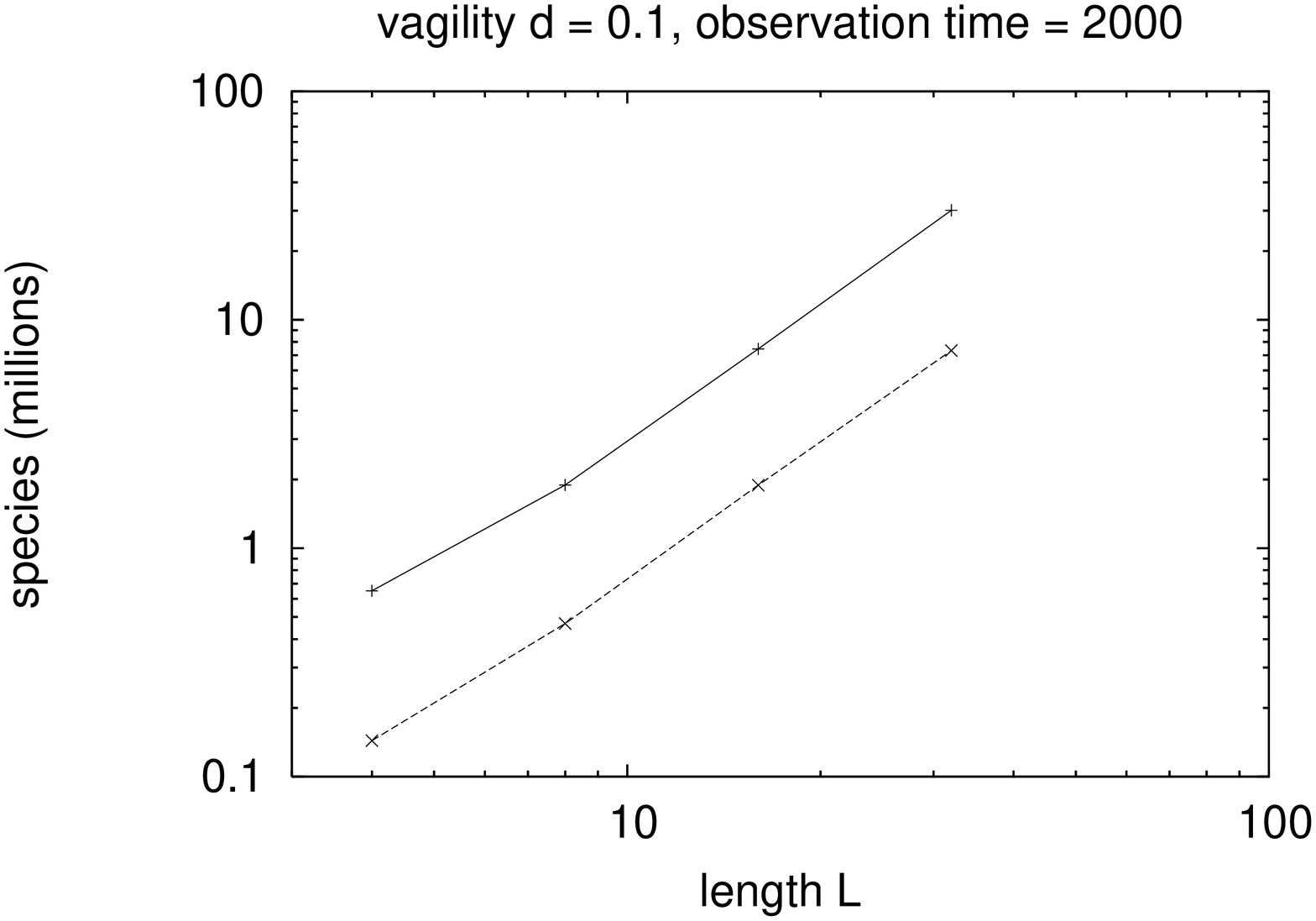}
\includegraphics[angle=-90,scale=0.25]{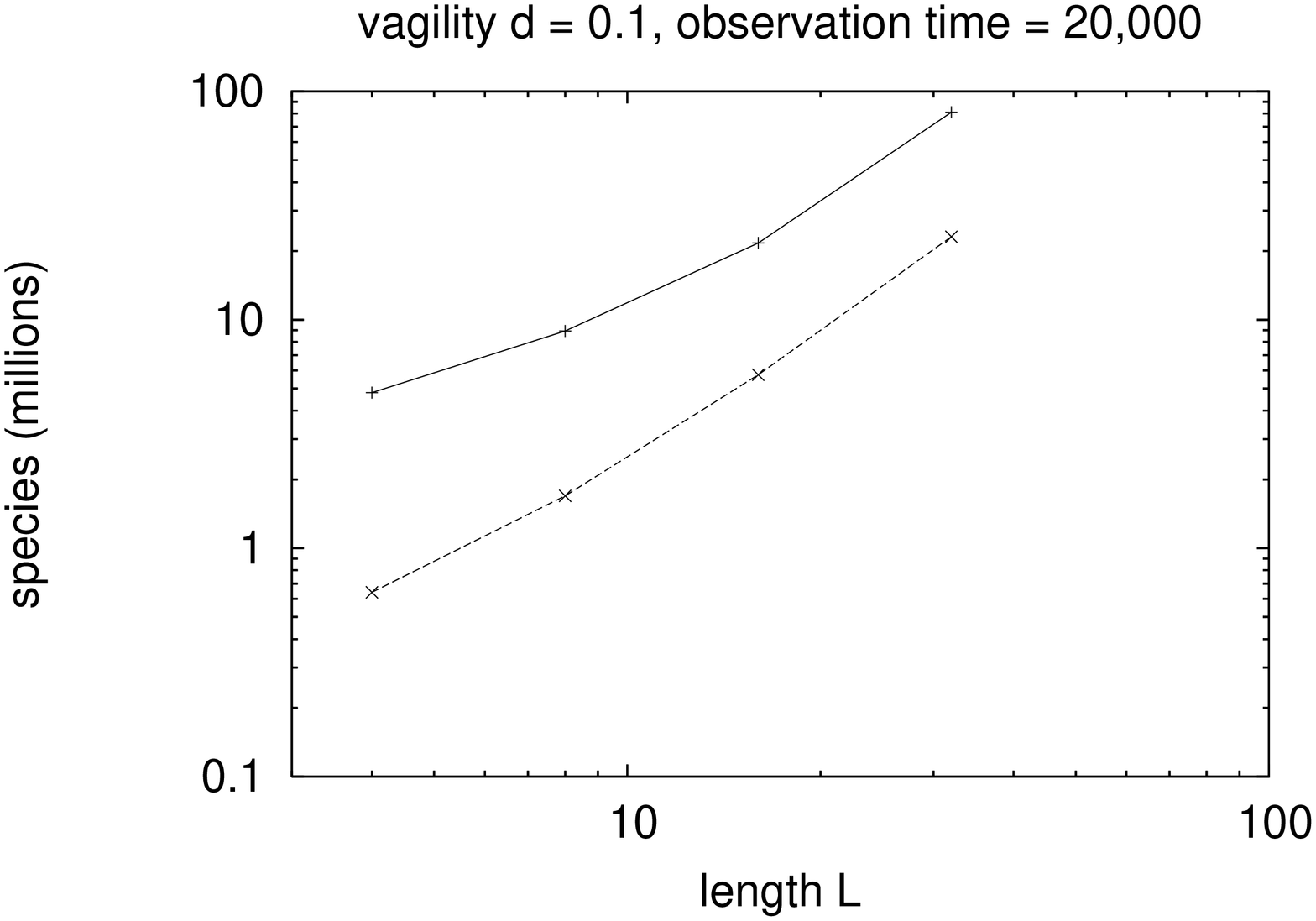}
\includegraphics[angle=-90,scale=0.25]{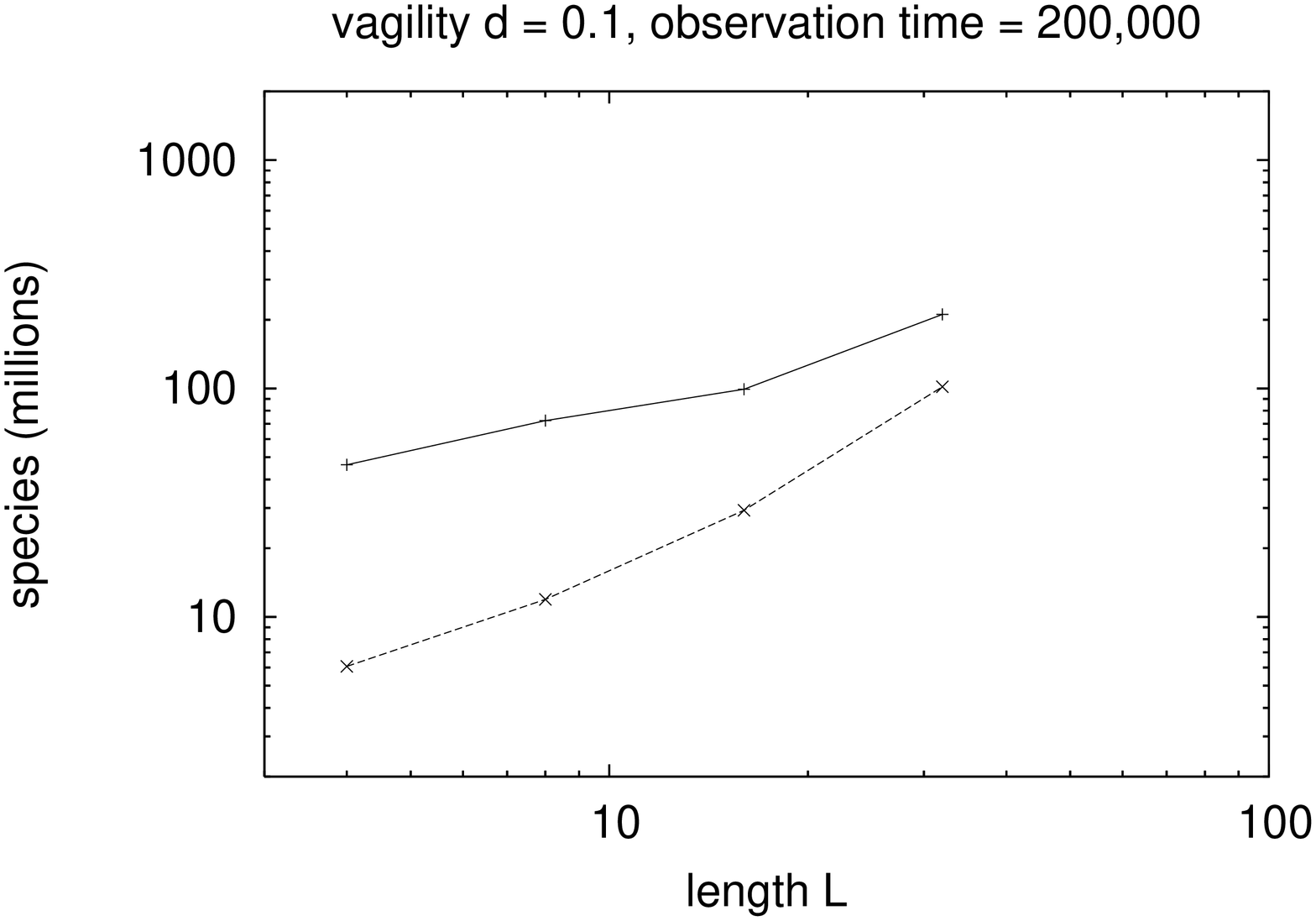}
\includegraphics[angle=-90,scale=0.25]{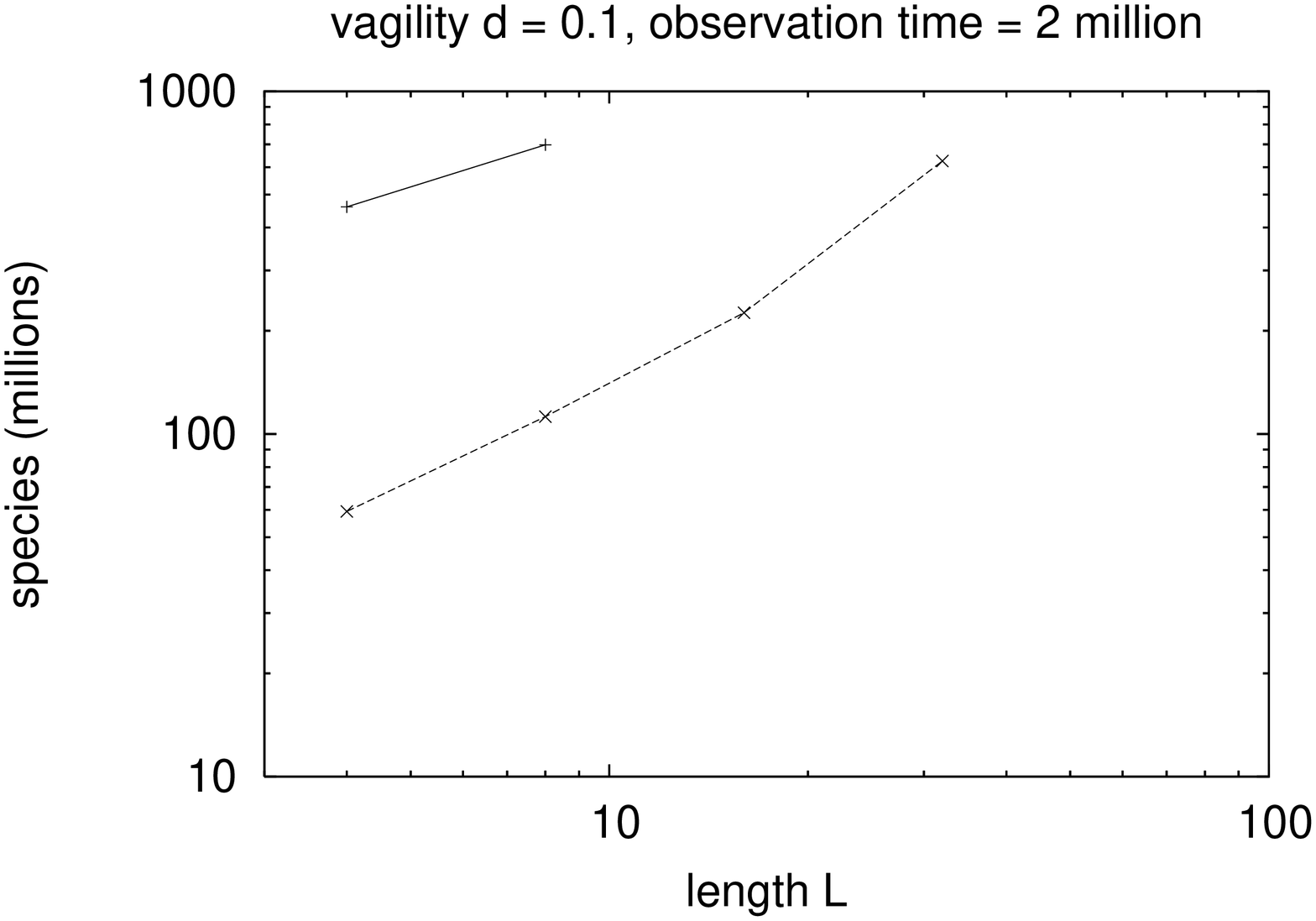}
\end{center}
\caption{Variation of the number $N$ of species with the length $L$ of 
the square, $L = 4$, 8, 16 and 32.  The vagility is $d=0.1$ for all four
cases. Upper lines = tropics, lower lines = polar.
}
\end{figure}

Fig.1 sums up the species number $N$ 
over all lattice sites and over all time steps after 
equilibration. The headlines give the observation times
varying from two thousand to two million time steps, for
various lattice sizes.

We see that for short times the species barely had a chance to 
move much from their site of origin, and thus $N$ is roughly
proportional to the area: $D = 2$. The longer the observation 
time is, the more could the species spread over the lattice,
and the smaller is the slope of our log-log plots.
It appears that the slope is smaller
in the tropics, which means that habitats are not narrower but somewhat larger
there than in polar regions, if they had sufficient time to spread. 

Fig.2 shows for a fixed observation time of 200,000 that the
slope $D$ becomes the larger the smaller the vagility is, thus
explaining the results of Stauffer and P\c ekalski (2005). 
For the smallest $d=0.001$ the data follow nearly
perfectly a line with slope $D = 2$, for the largest $d=0.1$ the curves starts 
with $D = 1$. For small $d$ one no longer sees the difference in
the polar and tropical slopes which is seen for large $d$. 

All these slopes $D$ agree with reality (Rosenzweig 1995) 
but do not come from good straight lines; our log-log plots
in general show upward curvature, and the slopes are those for intermediate
lattice sizes. Asymptotically for longer times and much larger lattice 
sizes $L$ we
expect the trivial proportionality with $D=2$ since then the range $\ell$
over which a species is spread obeys $ 1 \ll \ell \ll L$. The real Earth, 
however, may not correpond to these mathematical limits but to finite
sizes $L$ at finite times.

Fig.3 shows that the results are not merely a function of the product of 
vagility and observation time; varying $d$ influences many other properties 
(Rohde and Stauffer 2005) and not only the time scale.

\begin{figure}[hbt]
\begin{center}
\includegraphics[angle=-90,scale=0.31]{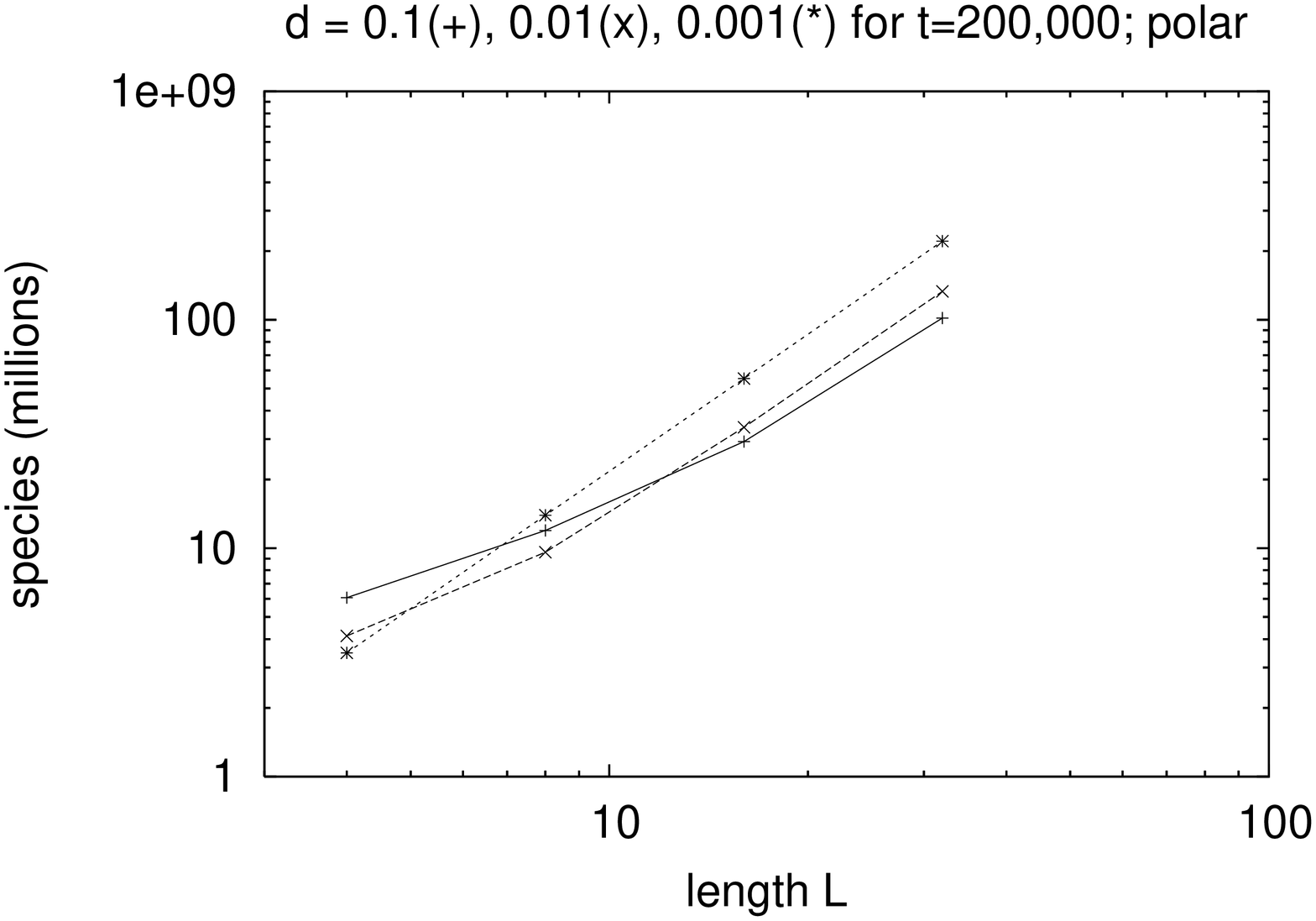}
\includegraphics[angle=-90,scale=0.31]{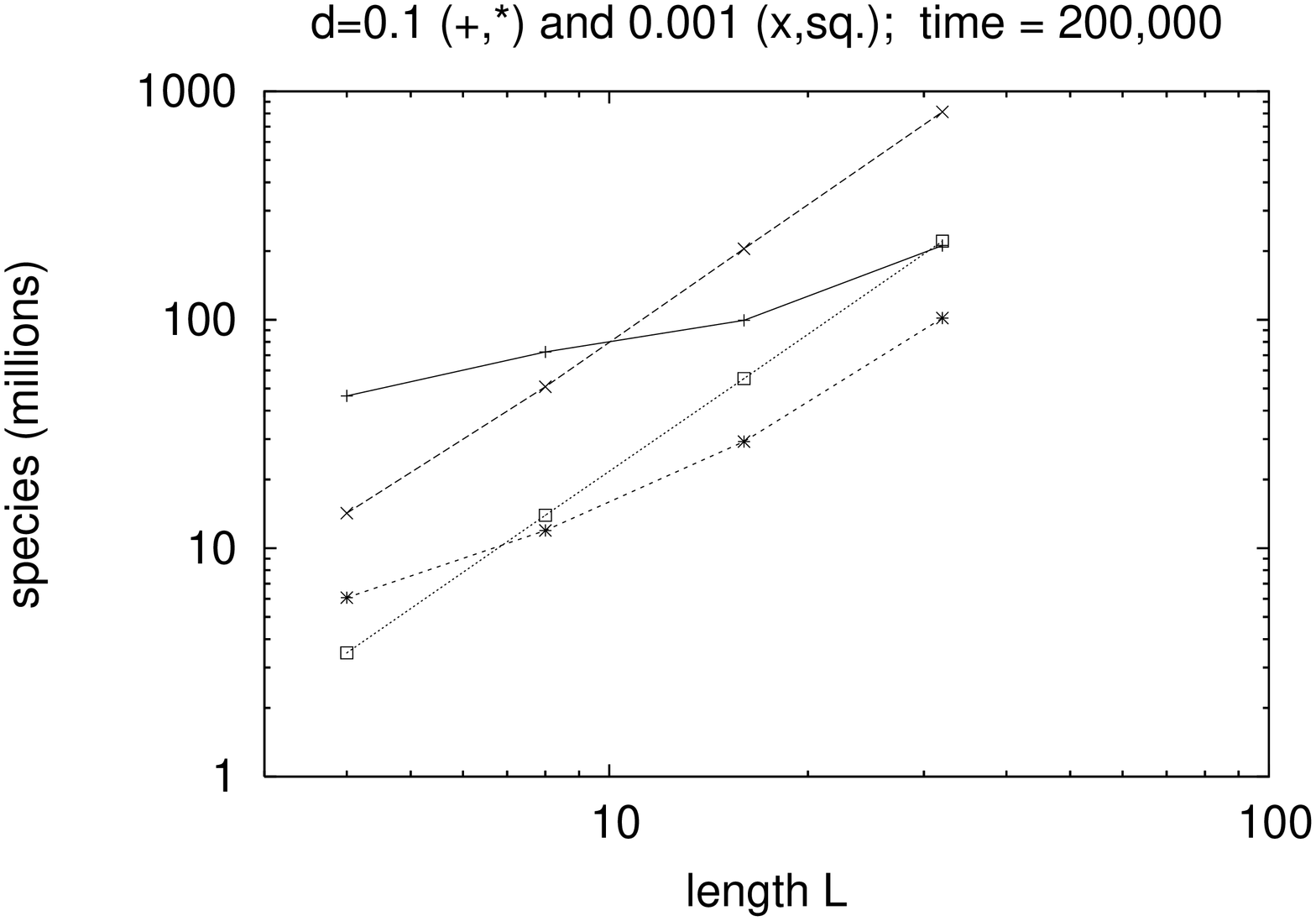}
\end{center}
\caption{Variation of $N$ versus $L$ for various $d$ at fixed observation time 
of 0.2 million. Upper part: polar; lower part: x and + for tropics, stars and 
squares for polar. 
}
\end{figure}

\begin{figure}[hbt]
\begin{center}
\includegraphics[angle=-90,scale=0.5]{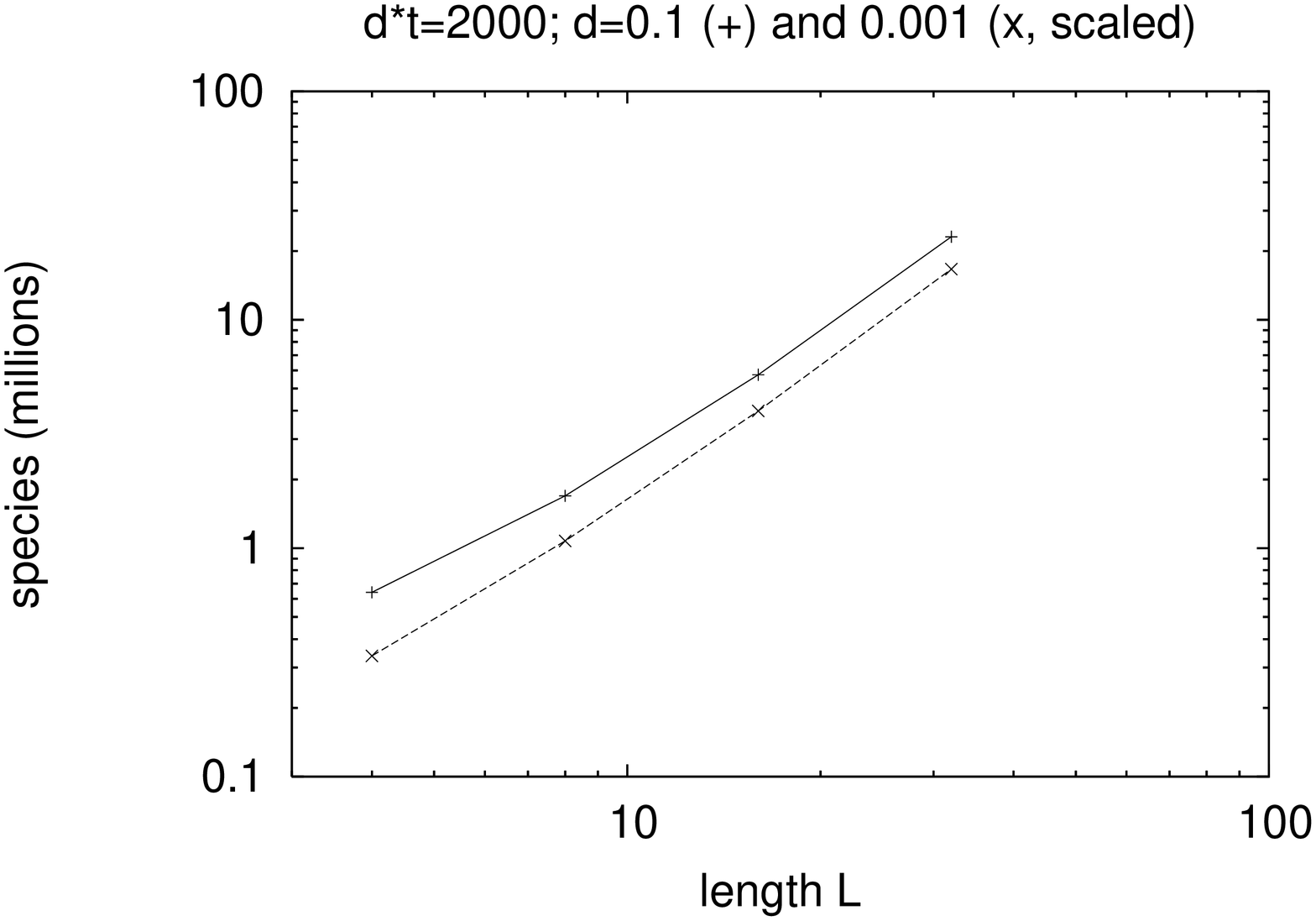}
\end{center}
\caption{Variation of polar $N$ versus $L$ for two vagilities $d$ and two 
observation times $t$ such that $dt$ is constant.
}
\end{figure}

\section{Discussion}

The findings presented in Fig. 1 contradict the latitude-niche width
hypothesis, for the niche dimension ``habitat width'',
according to which habitats are narrower in the tropics.
Indeed, they provide evidence for an opposite effect: habitats
are even larger near the equator than
at high latitudes. This agrees with the findings of Stauffer \& Rohde
(2006) who did not only fail to find support for Rapoport's rule, but
showed that latitudinal ranges are wider in the tropics, in agreement
with much empirical evidence.

The findings presented in Fig. 2 show that habitats are smaller in
species with little vagility, in accordance with the hypothesis,
developed in the context of Rapoport's rule, that young species (or
subspecies) with little vagility, which have not had sufficient time
to spread into wider areas, have narrower latitudinal ranges at low latitudes
(Rohde 1998).

Empirical evidence for the latitude-niche breadth hypothesis is
ambiguous. For example, Moore (1972) found that the average tropical
species occupies about half as much of the intertidal zone as the
average temperate species. According to MacArthur (1965, 1969),
tropical species often have a spottier distribution than
high-latitude ones.  Concerning one aspect of the habitat of animals
and plants, i.e. their latitudinal ranges, Stevens (1989) provided
evidence that some plant and animal species have narrower latitudinal
ranges in the tropics, referring to this phenomenon as Rapoport's
rule. Some of the numerous subsequent studies also provided evidence
for the rule (review in Rohde 1999).

However, support for the existence of narrower habitats in the tropics
is far from unequivocal. The studies that did not find support for
Rapoport's rule are more numerous than those that did, and in those
cases in which species have larger latitudinal ranges at high
latitudes, the increase is often restricted to high latitudes above
approximately 40-50 $^o$ N and S (review in Rohde 1999). Rohde (1996)
therefore suggested that the rule describes a local phenomenon, the
result of the extinction of species with narrow ranges during the ice
ages.

1) Several authors (e.g. Beaver 1979; review in Novotny \& Basset 2005)
have studied possible differences in host specificity of herbivorous
insects in tropical and temperate climates. No major differences were
found. 2) Detailed studies deal with latitudinal gradients in habitat
width of parasites of marine fish. Rohde (1978) has shown that host
ranges (the number of host species infected) of ectoparasitic
Monogenea infecting the gills are more or less the same at all
latitudes, whereas host ranges of another group of (endoparasitic)
flatworms, the Digenea, are markedly greater at high latitudes.
However, when correction was made for intensity and prevalence of
infection, host specificity was the same and very high at all
latitudes for both groups (Rohde 1980). Other niche dimensions of
these parasites, such as geographical range and microhabitat width,
were also examined and found not to be correlated with diversity,
although the data sets were small and more studies are needed. Host
size may on average be smaller in the tropics, due to the very large
number of host species, many of them small (Rohde 1989). 3)
Lappalainen and Soininen (2006) analysed the determinants of fish
distribution and the variability in species' habitat breadth and
position along latitudinal gradient of boreal lakes and found that
the regional occupancy of species was more strongly governed by the
habitat position than the habitat breadth. The cool water species
(percids and cyprinids) showed significant decrease in habitat breadth
towards higher latitudes (and not towards lower latitudes, expected
by the latitude-niche breadth hypothesis). 4)  Some further examples
are discussed in V\'azquez \& Stevens (2004).

V\'azquez \& Stevens (2004) have reviewed the evidence for the
latitude-niche breadth hypothesis, using meta-analytical techniques.
They found that the results of the meta-analysis do not permit
rejection of the null hypothesis of there being no correlation
between latitude and niche breadth. They also critically examined the
two assumptions on which MacArthur's hypothesis are based, i.e., 1)
that there is a latitudinal gradient in population variability, and
2) that there is a relationship between population variability and
niche breadth. These assumptions are widely accepted (e.g., May
1973). They claim that the tropics have greater stability and less
seasonality than temperate regions, making populations more stable,
thus allowing narrower niches. However, Rohde (1992) has pointed out
that there may be extreme variations in temperature, salinity and
currents in tropical shallow waters, such as high diversity coral
reefs. Such variations may occur over short time spans of a few
hours. The meta-analysis of V\'azquez \& Stevens (2004) shows that
available evidence does not support the view of an increasing
population variability with latitude, and evidence for narrower
niches of less variable populations is at best equivocal and does not
permit rejection of the null hypothesis of no relationship.

In spite of these criticisms of the mechanism involved, there could
be a latitudinal gradient in niche width due to other mechanisms.
V\'azquez \& Stevens (2004) suggest such a mechanism. Greater
specialization may be a by-product of the latitudinal gradient in
species diversity, because nestedness leads to an asymmetric, i.e.
faster increase of specialized species than of communities. In other
words, nestedness and asymmetric spezialisation tend to increase with
the number of species in a network. V\'azquez \& Stevens (2004) pay
particular attention to parasites. Nestedness of interactions between
species has, for example, been observed in marine Monogenea (Morand
et al. 2002), for which group, however, host specificity does not
change with latitude. Overall, nestedness is not common among
parasites of fish (Rohde et al. 1998; Poulin \& Valtonen 2001).

Finally and importantly, the latitude-niche breadth hypothesis as
formulated by MacArthur and his followers makes equilibrium
assumptions, and it implicitly and explicitly assumes that habitat
space is more or less filled with species. However, there is much
evidence that there is an overabundance of vacant habitats and that
most ecological systems are far from saturation (for a discussion and
examples see Rohde 2005). This removes the very basis on which the
hypothesis rests. The Chowdhury model does not make equilibrium
assumptions and incorporates vacant niches. Our simulations using
this model are further evidence against the latitude-niche breadth
hypothesis: tropical vagile species that have had sufficient time to
spread away from their original habitat, do not have narrower but
wider habitats than high latitude species.

How can we reconcile our results, that habitats of species are
somewhat larger in the tropics than at higher latitudes, with the
well known latitudinal gradient in species diversity? One possible
explanation is the idea of Terborgh (1973) and Rosenzweig (1995),
that tropical zones are generally larger and therefore stimulate
speciation and inhibit extinction. That larger areas (all other
conditions being equal or at least similar) often accommodate more
species, is well established. For example, at the level of
geographical area, Blackburn and Gaston (1997) found that there is
indeed a relationship between the land area and species richness of a
region once tropical species are excluded. This relationship is
independent of the latitude and productivity of regions.  A study on
South American mammals (Ruggiero 1999) confirmed this: The number of
principally extra-tropical mammal species per unit area depends on
the biome area (for further examples see Rosenzweig 1995). However,
as pointed out by Rohde (1998), although area matters, it cannot be
the primary cause of the latitudinal diversity gradients: many high
diversity tropical areas are much smaller than low diversity areas at
high latitudes. Many recent studies have provided support for this
view (e.g., MacPherson 2002: area size does not explain the
latitudinal pattern in benthic species richness on a large spatial
scale. Willig \& Bloch 2006: "area does not drive the latitudinal
gradient of bat species richness in the New World. In fact, area
represents a source of noise rather than a dominant signal at the
focal scale of biome types and provinces in the Western Hemisphere").
- Our results, that the habitats occupied by species are somewhat
larger in the tropics than at higher latitudes, mean, with regard to
latitudinal gradients, that there must be much overlap between
habitats, leading to a far greater diversity in tropical than in high
latitude areas of the same size. The larger (compared with high
latitude) tropical areas (in Africa and the IndoPacific) would
aggravate this. The overlap postulated here resembles the "Rapoport
rescue effect" of Stevens (1989), according to which tropical species
frequently "spill out" from their preferred habitat into adjacent
less favourable ones, thus explaining the high diversity there.
However, it is not necessary to distinguish favourable and less
favourable habitats: species may simply "spill out" from the habitat where 
they have originated, into adjacent habitats that are as suitable.

\section{Appendix: Model details}

The Chowdhury model of ecosystem has been described and modified in many
publications since 2003, and we give here only a short outline. Individuals
are born, mature, produce offspring asexually, and die with a probability 
increasing exponentially with age after maturity. At most 100 animals fit 
into one habitat. Six trophic levels define pre-predator relations: The upper 
levels feed on the adjacent lower ones. The topmost level has one habitat, the
second two, then 4, 8, .... At each iteration, with one percent probability
the food habits, minimum age of reproduction, and number of births per
iteration mutate randomly, allowing self-organisation of these parameters
through selection of the fittest. Death may come from from being eaten by a 
predator, from starvation, or from old age (with a high lifespan on the top 
food levels and a low lifespan on the bottom levels). If a species becomes 
extinct, then with probability 0.0001 per iteration the empty habitat is filled
by another species. Each of the $L^2$ lattice sites carries such an ecosystem,
each with dozens of living species. At the beginning, each different species 
gets a different number as its name. The number of differents species first 
decays with time (=iterations) and then fluctuates about some low average value.

Then with probability $d$ at each iterations
a species can migrate into a randomly selected
neighbour site, if the corresponding habitat on that neighbour site is empty at 
that time. A random fraction of the population moves, the rest stays at the 
old site. Both parts of the population carry the same name, and in this way
are counted as only once species spreading over more than one site. Summing
up over all different surviving names we obtain the number of different
species at that moment.
\bigskip

{\bf References}

\parindent 0pt
\medskip
Beaver, R.A. (1979). Host specificity of temperate and tropical
animals. Nature 281, 1139-141.

\medskip
Billari, F.C., Fent, T., Prskawetz, A. \& Scheffran, J. (Eds.) (2006).
Agent-based computational modelling, Heidelberg: Physica-Verlag .

\medskip
Blackburn, T.M. \& Gaston, K.J. (1997). The relationship between
geographic area and the latitudinal gradient in species richness in
New World birds. Evolutionary ecology 11, 195-204.

\medskip
Chowdhury, D. and Stauffer, D. (2005). 
Evolutionary ecology in silico: Does physics help in understanding
the ``generic'' trends ?  J. Biosci. (India) 30, 277-287.

\medskip
Grimm, V. \& Railsback, S.F. (2005). Individual-based modeling and
ecology. Princeton University Press, Princeton.


\medskip
Lappalainen, J. \& Soininen, J. (2006). Latitudinal gradients in niche
breadth and position - regional patterns in freshwater fish.
Naturwissenschaften 93, 246-250.

\medskip
MacArthur, R.H. (1965). Patterns of species diversity. Biological
Reviews 40, 510-533.

\medskip
MacArthur, R.H. (1969). Patterns of communities in the tropics.
Biological Journal of the Linnaean Society 1, 19-30.

\medskip
MacArthur, R.H. (1972). Geographical Ecology. Princeton University
Press, Princeton, NJ.

\medskip
MacArthur, R.H. \& Wilson E.O. (1967). An equilibrium theory of
insular zoogeography. Evolution 17, 373-387.

\medskip
MacPherson, E. (2002). Large-scale species-richness gradients in the
Atlantic Ocean. Proceedings of the Royal Society of London,
B-Biological Sciences 269, 1715-1720.

\medskip
May, R.M. (1973). Stability and complexity in model ecosystems.
Princeton University Press, Princeton N.J.

\medskip
Moore, H. B. (1972). Aspects of stress in the tropical marine
environment. Advances in marine Biology 10, 217-269.

\medskip
Morand, S., Rohde, K. \& Hayward, C.J.  (2002). Order in parasite
communities of marine fish is explained by epidemiological processes.
Parasitology, 124, S57-S63.

\medskip
Novotny, V. \& Basset, Y. (2005). Host specificity of insect
herbivores in tropical forests. Proceedings of the Royal Society B -
Biological Sciences 272, 1083-1090.

\medskip
P\c ekalski, A. (2004). A short guide to predator and prey lattice models by
physicists. Computing in Science and Engineering 6, 62-66 (Jan/Feb 2004).

\medskip
Poulin, R. \& Valtonen, E. T.  (2001). Nested assemblages resulting
from host size variation: the case of endoparasite communities in
fish hosts. International Journal for Parasitology 31, 1194-1204.

\medskip
Rohde, K. (1978). Latitudinal differences in host-specificity of
marine Monogenea and Digenea. Marine Biology 47, 125-134.

\medskip
Rohde, K. (1980). Host specificity indices of parasites and their
application. Experientia 36, 1370-1371.

\medskip
Rohde, K. (1989). Simple ecological systems, simple solutions to
complex problems? Evolutionary Theory 8, 305-350.

\medskip
Rohde, K. (1992). Latitudinal gradients in species diversity: the
search for the primary cause. Oikos 65, 514-527.

\medskip
Rohde, K. (1996). Rapoport's Rule is a local phenomenon and cannot
explain latitudinal gradients in species diversity. Biodiversity
Letters 3, 10-13.

\medskip
Rohde, K. (1998). Latitudinal gradients in species diversity; area
matters, but how much? Oikos 82, 184-190.

\medskip
Rohde, K. (1999). Latitudinal gradients in species diversity and
Rapoport's rule revisited: a review of recent work, and what can
parasites teach us about the causes of the gradients? Ecography 22,
593-613 (invited Minireview on the occasion of the 50th anniversary
of the Nordic Ecological Society Oikos). Also published in Fenchel,
T. ed.: Ecology 1999-and tomorrow, pp. 73-93. Oikos Editorial Office,
University Lund, Sweden.

\medskip
Rohde, K. (2005). Nonequilibrium ecology. Cambridge University Press,
Cambridge.

\medskip
Rohde, K., Heap, M. \& Heap, D. (1993). Rapoport's rule does not
apply to marine teleosts and cannot explain latitudinal gradients in
species richness.  American Naturalist, 142, 1-16.

\medskip
Rohde, K., Worthen, W., Heap, M., Hugueny, B. \& Guégan, J.-F. (1998).
Nestedness in assemblages of metazoan ecto- and endoparasites of
marine fish. International Journal for Parasitology, 28, 543-549.

\medskip
Rohde, K. \& Stauffer, D. (2005). Simulations of geographical trends
in the Chowdhury ecosystem model. Advances in Complex Systems 8,
451-464.

\medskip
Rosenzweig, M. L. (1995). Species diversity in space and time.
Cambridge University Press, Cambridge.

\medskip
Rosenzweig, M.L. \& Ziv, Y. (1999). The echo pattern of species
diversity: patterns and processes. Ecography 22, 614-629. 

\medskip
Ruggiero, A. (1999). Spatial patterns in the diversity of mammal
species: A test of the geographic area hypothesis in South America.
EcoScience 6, 338-354.

\medskip
Stauffer, D \& P\c ekalski, A. (2005) Species numbers versus area in Chowdhury 
Ecosystems. www.arXiv.org: qbio.PE/0510021.

\medskip
Stauffer, D. \& Rohde, K. (2006). Simulation of Rapoport's rule for
latitudinal species spread. Theory in Biosciences 125, 55-65.

\medskip
Stauffer, D., Kunwar, A. \& D. Chowdhury,D. (2005).
Evolutionary ecology in-silico: evolving foodwebs, migrating population
and speciation. Physica A 352, 202-215 (2005).

\medskip
Stauffer, D., Moss de Oliveira, S., de Oliveira, P.M.C., \& S\'a
Martins, J.S. (2006). Biology, Sociology, Geology by Computational
Physicists. Amsterdam, Elsevier.

\medskip
Stevens, G.C. (1989). The latitudinal gradients in geographical
range: how so many species co-exist in the tropics. American
Naturalist 133, 240-256.

\medskip
Terborgh, J. (1973). On the notion of favourableness in plant
ecology. American Naturalist 107, 481-501.

\medskip
V\'azquez, D.P. \& Stevens, R.D. (2004). The latitudinal gradient in
niche breadth: concepts and evidence. American Naturalist 164,
E1-E19.

\medskip
Willig, M.R. \& Bloch, C.P. (2006). Latitudinal gradients of species
richness: a test of the geographic area hypothesis at two ecological
scales. Oikos 112, 163-173.

\end{document}